\documentstyle[12pt,aaspp4]{article}

\def\etal{{et al.}}
\def\asca{{\it ASCA}}

\def\ros{{\it ROSAT}}

\newbox\grsign \setbox\grsign=\hbox{$>$} 
\newdimen\grdimen \grdimen=\ht\grsign
\newbox\simlessbox \newbox\simgreatbox \newbox\simpropbox
\setbox\simgreatbox=\hbox{\raise.5ex\hbox{$>$}\llap
     {\lower.5ex\hbox{$\sim$}}}\ht1=\grdimen\dp1=0pt
\setbox\simlessbox=\hbox{\raise.5ex\hbox{$<$}\llap
     {\lower.5ex\hbox{$\sim$}}}\ht2=\grdimen\dp2=0pt
\setbox\simpropbox=\hbox{\raise.5ex\hbox{$\propto$}\llap
     {\lower.5ex\hbox{$\sim$}}}\ht2=\grdimen\dp2=0pt

\begin{document}

\title{X-ray Observations of the Seyfert galaxy LB~1727 (1H 0419-577)} 

\author {T.J.Turner\altaffilmark{1,2,3}, 
I.M. George \altaffilmark{1, 2}, D. Grupe \altaffilmark{4}, 
K. Nandra \altaffilmark{1, 5}, 
R.A. Remillard \altaffilmark{6}, \\
K.M. Leighly \altaffilmark{7},    
H.L. Marshall \altaffilmark{8},
S.B. Kraemer \altaffilmark{9},
D.M. Crenshaw \altaffilmark{9} 
 } 

\altaffiltext{1}{Laboratory for High Energy Astrophysics, Code 660,
	NASA/Goddard Space Flight Center,
  	Greenbelt, MD 20771}
\altaffiltext{2}{Universities Space Research Association}
\altaffiltext{3}{Present address, University of Maryland, Baltimore County}
\altaffiltext{4}{MPE, Giessenbachstr., D-85748 Garching, Fed. Rep. Germany}
\altaffiltext{5}{NAS/NRC Research Associate}
\altaffiltext{6}{Massachusetts Institute of Technology, Cambridge, MA 02139}
\altaffiltext{7}{Columbia Astrophysics Laboratory, Columbia University, 538 
  West 120 St., New York, NY 10027}
\altaffiltext{8}{Eureka Scientific, Inc.}
\altaffiltext{9}{Catholic University of America, Code 681, 
	NASA/Goddard Space Flight Center,
  	Greenbelt, MD 20771}
	
\begin{abstract}

We discuss the properties of the Seyfert 1.5 galaxy 
LB~1727, also known as 1H~0419-577, from 
X-ray observations obtained by \asca\ and \ros\ along with 
optical observations from earlier epochs. 
The source flux was 
$F_{2-10} \simeq 10^{-11}\ {\rm erg\ cm^{-2}\ s^{-1}}$ during the {\it ASCA}
observations which were carried out 1996 Jul --  Aug, and we find 
only modest ($\lesssim 20$\%) variations in the flux in this band, within or 
between these observations. 
In contrast, a daily monitoring campaign 
over 1996 Jun -- Sept by the \ros\ HRI instrument reveals the 
soft X-ray (0.1-2~keV) flux to have increased 
by a factor $\simeq 3$. 
Significant variations were also observed down to 
timescales of $\sim 40$ ks.

We find the 2 -- 10~keV continuum can be 
parameterized as a power-law with a photon index 
$\Gamma \sim 1.45-1.68$ across $\sim$ 0.7 -- 11 keV in the rest-frame.
We also report the first detection of iron K$\alpha$ line emission 
in this source. Simultaneous {\it ASCA} and {\it ROSAT} data show 
the X-ray spectrum to steepen sharply at a rest-energy  $\sim 0.75$ keV, 
the spectrum below this energy can be parameterized as a 
 power-law of slope $\Gamma \sim 3.6$. 
The X-ray emission appears to be unattenuated and we find 
that ionized gas alone cannot produce such a sharp spectral break. 
Even allowing the presence of such gas, the 
simultaneous {\it ASCA} and HRI data demonstrate that the 
underlying continuum is required to steepen below $\sim 0.75$ keV. 
Thus LB~1727 is one of the few Seyferts for which we can rule out 
the possibility that the presence of a warm absorber is {\it solely}
 responsible for the spectral steepening in the soft X-ray regime. 

Consideration of the overall spectral-energy-distribution for this source 
indicates the presence of a pronounced XUV-bump visible in optical, 
ultraviolet and soft X-ray data. The source appears relatively weak in 
infrared 
emission and so if dust exists in the source, it is not excited by 
the nuclear radiation.

\end{abstract}


\section{Introduction}
\label{sec:intro} 

It has long been suggested that the putative accretion disk around the black
hole of active galactic nuclei (AGN) will emit 
copious amounts of UV and soft X-ray 
radiation (Shields 1978, Rees 1984). Thus
the spectral energy distribution (SED) of an AGN across this range 
should provide crucial information about the accretion disk, 
reprocessing mechanisms close to the active nucleus and 
physical conditions in the circumnuclear material.

Determination of the X-ray to ultraviolet (XUV) continuum in AGN has been
extremely difficult because of the severe attenuation of photons of these
energies by even small amounts of material along the line-of-sight to the AGN.
However, some indication of the strength of the unseen continuum has been
inferred from the strengths of emission lines such as He{\sc ii} $\lambda
1640$ (e.g. Mathews \& Ferland 1987). In fact, a long-standing suggestion has
been that there is a so-called ``blue-bump'' of continuum emission, peaking in
the unseen XUV regime (Shields 1978, Malkan \& Sargent 1982). The advent of
new and sensitive satellites has started to narrow the ``unobservable''
bandpass. For example, observations of high-redshift quasars by {\it HST} has
enabled measurement of the continuum up to the Lyman limit, while the {\it
ROSAT} Wide Field Camera  and the Extreme Ultraviolet Explorer {\it EUVE} have
provided measurements in the soft X-ray band ($\sim 0.1-1$ keV).
Four Seyferts are listed as detections by the WFC (Pounds \etal\ 1993), 
furthermore, (at least) 17 Seyferts were detected by {\it EUVE} 
(Marshall \etal\ 1995, Fruscione 1996). 
Rapid variability in the {\it EUVE} flux 
of several sources indicated the observed radiation to be originating in the
inner nucleus (e.g. Marshall \etal\ 1996; Marshall \etal\ 1997).  

Recent work by Zheng \etal\ (1997, 1995) has suggested the form of the unseen
XUV spectrum is falling between the Lyman limit (at 912 \AA) and 
$\sim 0.5 $ keV as $f_v \propto v^{-2}$. 
Laor \etal\ (1997) combine this with a mean soft X-ray
spectrum, based upon {\it ROSAT} observations of quasars, to dispute the
existence of a large XUV-bump. Korista, Ferland and Baldwin (1997) have
discussed the problem that extrapolating the known soft X-ray spectrum of AGN,
there appear to be too few 54.4 eV photons to account for the strength of the
observed He{\sc ii} lines. They consider the possibility that the broad-line
clouds see a harder continuum than the observer does, or that the XUV spectrum
has a double-peaked shape. 
Clearly, a careful determination of the detailed shape of the
XUV continuum would provide a large step forward in our understanding of many
fundamental processes in AGN. 
While numerous observations of AGN have been performed in 
the past, little is known about the spectral shape below 
$\sim 0.6 $ keV. Previous soft X-ray observations have yielded 
narrow-band spectra with low energy resolution (e.g. {\it ROSAT}, 
{\it Einstein} IPC, {\it EXOSAT} CMA). {\it ASCA} 
has only yielded reliable spectra above 0.6 keV (although data may 
eventually be reliable down to 0.4 keV, with improved calibration). 
Disagreements between {\it ASCA} and {\it ROSAT} data,
in the overlap bandpass (0.6-2 keV) have led 
to some controversy as to whether 
the underlying continua of emission-line AGN actually do 
steepen to softer X-ray energies or whether there is a 
miscalibration between instruments. {\it ASCA} and {\it ROSAT} data aside, 
some sources definitely show a significant spectral softening below 
$\sim 1$ keV in data from earlier missions (e.g. Arnaud 1985, Turner \etal\ 
1993), however, in most cases, the physical origin of this effect is
ambiguous. It is now known that the majority of Seyfert~1 galaxies suffer 
attenuation by ionized material (Reynolds 1997; 
George \etal\ 1998) which can have  a reduced opacity 
in the soft X-ray band and produce an observed spectrum 
which steepens to soft energies.  In most cases studied to date, X-ray 
spectra do not allow us to distinguish between 
a single continuum component attenuated by ionized material, and 
steepening of the underlying emission spectrum. 
Clearly, the best sources for attempting to unambiguously distinguish 
between the aforementioned pictures should be 
bright in the soft X-ray regime, and have minimal absorption.
LB~1727 is the fourth  brightest Seyfert galaxy detected by {\it EUVE} 
(Marshall \etal\ 1995), yet has a very flat spectrum in the hard X-ray 
regime (Guainazzi \etal\ 1998) compared to most Seyfert~1 galaxies 
(George \etal\ 1998 and references therein). These two properties suggest  
the source might show 
a marked steepening of spectral slope in the soft X-ray regime; this, combined 
with the low Galactic column along the line-of-sight, makes LB~1727 a 
good target to search for the soft X-ray tail of an XUV-bump. 

\section{The Seyfert galaxy LB~1727}

LB~1727 is a 
very blue object first noted by Luyten \& Miller (1956) during
their search for faint blue stars. The X-ray source 1H~0419-577 
was discovered in the sky-survey performed 
by the HEAO-1/A1 experiment, covering the 
0.25 -- 25 keV band,  during late 1977 to early 1978 (Wood 
\etal\ 1984). At that time, the 2 -- 10 keV flux was 
$\sim 2 \times 10^{-11} {\rm erg\ cm}^{-2} {\rm s}^{-1}$ in the rest-frame. 
The HEAO-1/A1 survey was followed by a program of optical identifications, 
and an analysis of HEAO-1/A3 X-ray positions identified a Seyfert
galaxy (V = 14.3; z = 0.104) as the optical counterpart
to the HEAO-1 X-ray source (Brissenden et al. 1989).
The source was later detected in the 
Einstein slew survey (Elvis \etal\ 1992) and by {\it ROSAT} with 
both the Position Sensitive Proportional 
Counter (PSPC) and the Wide Field Camera (WFC; Pounds \etal\ 1993). Thomas 
\etal\ (1998) confirmed LB~1727 to be a bright X-ray source and 
a Seyfert 1.5 galaxy (Grupe 1996; Thomas \etal\ 1998). 

Confusion has arisen on several ocassions, as to whether 
LB~1727 and 1H 0419-577 are the same source or not; they are. 
An error in some {\it EUVE} 
finding charts led to confusion because 
1H~0419-577 and LB~1727 were wrongly marked as 
two different sources in those charts. This error eventually resulted 
in an incorrect position  assignment for LB~1727, in the 
Veron-Cetty \& Veron catalog (1996). 
Despite this early confusion, Marshall \etal\ 1995 
did correctly identify 1H 0419-577 as a strong {\it EUVE} detection 
of an AGN. 
Comparison between the original position found by 
Luyten \& Miller (1956), and the optical, 
{\it ROSAT} and {\it EUVE} positions confirm LB~1727 and 
1H~0419-577 to be 
the same source, a Seyfert galaxy, and the origin of all data 
presented here. 

Guainazzi \etal\ (1998) report the results from 
a {\it BeppoSAX} observation of LB~1727 
performed 1996 September 30. The exposure time 
was $\sim 23$ ks with the Medium Energy Concentrator
Spectrometer (MECS) which has an effective bandpass $\sim 1.8 - 10$ keV. 
Unfortunately the 
Low Energy Concentrator Spectrometer (LECS), which covers the 0.1 -- 10 keV 
band, was switched off during the observation because of technical problems. 
The source was also marginally detected in the {\it BeppoSAX} 
Phoswich Detector System (PDS), providing some constraint on the flux up 
to 36 keV. 
The {\it BeppoSAX} data revealed a flat spectrum in the 1 -- 10 keV 
(observed) band, with photon index 
$\Gamma \sim 1.6$ and no evidence for line emission from the K-shell of iron. 
Guainazzi \etal\ (1998) also reported an index $\Gamma \sim 2.7$ from 
PSPC data covering the 0.1 -- 2 keV band, indicating marked spectral 
variability and/or a sharp steepening of the spectrum below $\sim 2$ keV. 

Here we present the results of new X-ray observations of LB~1727, along
with several optical spectra. In \S3 we present analysis of 
optical spectra from 1988 and 1993. In \S4 we detail the timing and spectral 
results from 
two {\it ASCA} observations from July and August 1996. 
In \S5 we show the {\it ROSAT} HRI results from a 
two-month  monitoring campaign. In \S6 
we discuss analysis of some simultaneous HRI and {\it ASCA} data, 
and constraints on the amount of ionized gas along the line-of-sight, 
and in \S7 we examine the PSPC data in the light of those results. 
In \S8 we discuss the overall 
SED of LB~1727, and compare it with other Seyfert galaxies. 
In \S9 we discuss all of these results 
along with implications regarding the presence of an 
XUV-bump. 

\section{The Optical Spectrum}
\label{sec:opt} 

LB~1727 has been observed several times in attempts to identify sources from
hard and soft X-ray surveys (Brissenden \etal\ 1989; Grupe 1996; Thomas \etal\
1998, Guainazzi \etal\ 1998). The optical position of the source is  (J2000)
$04^{\rm h}26^{\rm m}0.83^{\rm s}\  -57^{\rm o}12'0.45'' $, with an uncertainty 
of $1''$ radius.

The optical spectra  of LB ~1727 are shown in Fig.~\ref{fig:opt}. The 
first two panels 
show spectra accumulated during 1988 Feb 17 and 22, on the 
Australian National University (ANU) 2.3m telescope and the 
3.9m Anglo-Australian Telescope (AAT), respectively. 
The AAT spectrum was obtained with the combined use of the Image Photon
Counting System (3 \AA\ FWHM resolution) and the Faint Object Red Spectrograph
(20 \AA\ FWHM).  A dichroic mirror was used to split the incoming
beam at $\sim 5500$\AA\ to supply the photons for each device. Similarly,
the ANU observations were made with the Double Beam Spectrograph (3 \AA\ FWHM).
An optical spectrum was also accumulated 
1993 Sept 14 with the MPI/ESO 2.2m telescope at La Silla, using the 
Faint Object Spectrograph and Camera. 
The ESO 2.2m observations used grisms 
yielding 5 \AA\ FWHM resolution over the 6600 -- 7820 \AA\  and 
4640 -- 5950 \AA\ bandpasses,  plus 22 \AA\  FWHM over the 3400-9200 \AA\ 
bandpass; with 
exposure times of 15, 20 and 5 minutes respectively. The detailed reduction
method for the ESO 2.2m data is described in Grupe (1996, 1998b) and in 
Brissenden \etal\ (1989) for the ANU and AAT data.

Fig.~1 shows the continuum rises very steeply blueward of 4500 \AA, most visible in the 
AAT data, indicating the presence of a strong blue bump in this source.
The optical continuum has an energy index $\alpha_{opt}=0.01^+_-0.40$ 
measured between 4400 -- 7000 \AA\  (Grupe et al. 
1998a). The strong {\it EUVE} flux 
(Marshall \etal\ 1995) and steep spectrum in the soft X-ray regime 
(Guainazzi \etal\ 1998) might
lead us to expect to see an optical spectrum reminiscent of a 
Narrow Line Seyfert 1 galaxy (NLSy1); 
however, from an inspection of the line profiles 
in Fig.~1 it is clear that this is not the case. 
NLSy1s are those objects with 
the narrowest optical permitted lines in the distribution covered by Seyfert 1
galaxies. The widths of the broad components of 
H$\alpha$ and H$\beta$ are typically only slightly broader
than the forbidden lines. In general authors usually use 
FWHM H$\beta <2000$ km/s (or sometimes $< 1500$ km/s) and the 
presence of strong Fe{\sc ii} emission to distinguish 
between Seyfert 1s and NLSy1s. The ratio 
\verb+[+O{\sc iii}\verb+]+/H$\beta\ < 3$, is also used to differentiate 
NLSy1 from Seyfert~2 galaxies.  
In the case of LB~1727 the width of the broad H$\beta$
component does not fit into the NLSy1 definition. 
The AAT data yielded 
FWHM \verb+[+O{\sc iii}\verb+]+ $=580 ^+_-50 {\rm km\ s}^{-1}$ 
(1 $\sigma$ errors are quoted on optical measurements in this section) 
FWHM H$\beta_{narrow}=1225 ^+_-200 {\rm km\ s}^{-1}$ and 
H$\beta_{broad}=4200 ^+_-250 {\rm km\ s}^{-1}$. 
For the ANU data the  \verb+[+O{\sc iii}\verb+]+ line was fit with a 
gaussian profile, yielding  FWHM$=790 ^+_-30{\rm km\ s}^{-1}$. 
A template was made of this line profile, and this was 
scaled to the flux of the narrow 
component of H$\beta$. The scaled template was shifted to the wavelength 
of the narrow component of the H$\beta$ line and then subtracted from
the total H$\beta$ profile. The remaining H$\beta$ profile was then
dominated 
by the broad component of H$\beta$, which yielded a width measurement 
$2950 ^+_-100{\rm km\ s}^{-1}$. This left a residual narrow component of
H$\beta$ 
with FWHM=$1080 ^+_-100{\rm km\ s}^{-1}$. 
(It was not possible to satisfactorily apply this method to the AAT
data). The ESO 2.2m 
data yielded FWHM \verb+[+O{\sc iii}\verb+]+ $=450 ^+_-10 {\rm km\ s}^{-1}$ 
and H$\beta_{narrow}=700 ^+_-100{\rm km\ s}^{-1}$,
H$\beta_{broad}=2900 ^+_-100 {\rm km\ s}^{-1}$, corrected for instrumental 
resolution (Grupe \etal\ 1998b). 
The errors given in this section are 1$\sigma$. The measurements 
might imply variability in the width of H$\beta$, however, these data 
were all taken using different instrument combinations with 
a variety of spectral resolutions, and conclusive 
detection of variability could only be obtained by monitoring the source 
for several days with a single instrument. 

The \verb+[+O{\sc iii}\verb+]+/H$\beta$ 
ratio  ($\sim$ 10) is large and there is no evidence of 
Fe {\sc ii} emission,  which is most clearly evident from 
the AAT data. Grupe \etal\ (1998b) report 
an upper limit on the equivalent width for Fe {\sc ii} emission 
between 4250 - 5880 \AA~ to be 30\AA,  and an upper limit on the 
ratio Fe{\sc ii}/H$\beta$ of 0.4. 

To determine the Balmer decrement we used the total fluxes in the 
H$\alpha$ and H$\beta$ lines because it is not 
possible to estimate the broad and narrow decrements 
separately due to problems deconvolving the components of H$\alpha$. 
By using the \verb+[+O{\sc iii}\verb+]+ template we are able 
to subtract the narrow line emission from the H$\beta$ line, but this
is not possible with the H$\alpha$ line. 
Unfortunately, H$\alpha$ is also 
contaminated by the \verb+[+N{\sc ii}\verb+]+ 6548\AA~ and 6584\AA~ 
emission. To subtract the contributions
of these lines we subtracted 35\% of the 
\verb+[+O{\sc iii}\verb+]+5007 line flux from the
total H$\alpha$ line flux. 
(The \verb+[+N{\sc ii}\verb+]+/\verb+[+O{\sc iii}\verb+]+ ratio was 
suggested by Ferland \& Osterbrock, 1986).  Subtracting this 
contaminating flux allows us to make an estimate of the mean Balmer
decrement, H$\alpha$/H$\beta = 4.2$. 
This decrement can be interpreted as an optical 
extinction $A_V=1.21$ which in turn corresponds to an X-ray absorption 
$N_H \sim 1.8 \times 10^{21} {\rm cm}^{-2}$, assuming material with
a Galactic dust-to-gas ratio. However, the relatively low value of the 
decrement, and the uncertainty in that ratio 
makes this extinction determination somewhat tentative. 
In fact, the optical {\it continuum} is 
unattenuated. The latter might infer that the Balmer decrement is attributable 
to optical depth effects 
in the line-producing clouds rather than absorption by material in 
our line-of-sight. 

\section{ASCA Observations and Data Reduction}

{\it ASCA} (Makishima et al. 1996) has two solid-state 
imaging spectrometers 
(SISs; Burke \etal\ 1994) and two gas imaging spectrometers 
(GISs; Ohashi \etal\ 1996) sensitive across the 
$\sim$ 0.4 -- 10 keV and  $\sim$0.8 -- 10~keV bandpasses, respectively. 
LB~1727 was observed by {\it ASCA} 1996 July 22 -- 23 and  August 10 -- 11. 
The data were reduced in the same way as the Seyfert galaxies
presented in Nandra \etal\ (1997) and Turner \etal\ (1997). For details of 
the data reduction method see Nandra \etal\ (1997). 
For the July observation, data screening 
yielded effective exposure times of $\sim 24 $ ks in all four instruments. 
For the August observation the exposure times were $\sim 23$ ks in SIS and 
$\sim 26$ ks in the GIS instruments. 

\subsection{Time Variability}

During the July and August epochs observed by {\it ASCA}, 
the fluxes in the 0.5 -- 2 keV rest-frame 
(0.45 -- 1.81 keV observed-frame) 
were $4.8$ and $5.5 \times 10^{-12} {\rm erg\ cm}^{-2} {\rm s}^{-1}$, 
respectively, The 
2 -- 10 keV rest-frame fluxes (1.81 -- 9.06 keV observed-frame) 
were $0.94$ and $1.1 \times 10^{-11} {\rm erg\ cm}^{-2} {\rm s}^{-1}$. 
The flux variation between epochs is evident in the hard-band light curves 
(Fig.~\ref{fig:asca_lc}).  
The 2 -- 10 keV fluxes are consistent with that observed 
by {\it BeppoSAX} a month later on 1996 September 30 
(Guainazzi \etal\ 1998) and a factor of 2 lower than that observed 
by HEAO-1/A1 (Wood \etal\ 1984) 19 years earlier. The mean {\it ASCA} 
flux implies a luminosity of 
$L$(2-10 keV)$ = 4.8 - 5.6 \times 10^{44} {\rm erg\ s}^{-1}$ 
(assuming H$_{\rm 0}=50$, q$_{\rm 0}=0.5$).

We tested the light curves for variability using the $\chi^2$ statistic. 
The source showed no significant variability in either 
band when sampled in 256 s or 5760 s bins. 
Fig.~\ref{fig:asca_lc} shows the light curves constructed in the 0.5 -- 2 
keV and 
2 -- 10 keV observed bands and sampled on 5760 s, this integration time  
represents one {\it ASCA} orbit. There is 
a $\sim 60\%$ increase in soft flux  
and a $\sim 20\%$ increase in the 2 -- 10 keV flux between 
July and August 1996. 
Integration using 1024 s bins revealed evidence 
(at $> 90\%$ confidence) for flickering at the 
10 -- 30\% level in both bands during the July 
observation, but no significant variability during the August observation. 

\subsection{The X-ray Spectra} 

{\it ASCA} data cover the 0.4-10 keV band in the observers frame. 
However, the SIS data below an energy of 0.60 keV 
(0.66 keV rest-frame) were excluded from the spectral analysis 
as it is commonly accepted that there are uncertainties 
associated with the {\it ASCA} calibration in that band. 
Following George \etal\ (1998), we make use of the fact that 
the calibration uncertainty is considered to be $\lesssim 20$\%, and 
usually results in a systematic deficit of counts versus the predicted model. 
If data in the 0.40-0.60 keV band lie above 
the extrapolation of our model, 
then the source spectrum most likely does steepen in that 
band. Later  we take the approach of indicating 
where these data lie, although we never use them in a fit. 

\subsubsection{1996 July}

Data in the 4.5-6.8 keV range (5-7.5 keV in the rest-frame) were 
initially excluded, to remove temporarily the 
channels in which iron K$\alpha$ emission would occur if present in this 
source. This exclusion allows us to parameterize the 
continuum shape more easily. We first considered the continuum 
in the observed 1.81 -- 9.06 keV band (2-10 keV rest-frame)   
using a power-law attenuated by a 
column of neutral material. Preliminary fits showed no evidence for 
absorption, so the absorbing column density was fixed subsequently
 at the Galactic value estimated from 21 cm measurements, 
$N_H=2.25 \times 10^{20} {\rm cm}^{-2}$ (Dickey \& Lockman 1990). 
  It is worth noting that the extremely 
large {\it EUVE} count rate suggests the effective XUV absorbing-column 
may be even lower than this (cf Marshall \etal\ 1995). 
The 1996 July data yielded a photon index 
$\Gamma_{rest}$(2-10)$=1.48^+_-0.07$ and $\chi^2=280 $ 
for 319 degrees of freedom ($dof$). We then fit the same model 
to the full {\it ASCA} band, i.e. 
(0.6-10 keV observed-frame, 0.66-11.1 keV rest-frame, still 
excluding the iron K$\alpha$ band). This yielded 
$\Gamma_{rest}$(0.66-11.1)$=1.45^+_-0.03$ for 
$\chi^2=488/533\ dof$. Thus there 
is no evidence for significant spectral curvature down to a rest-energy 
of $\sim$ 0.7 keV. 
The ratio of the data compared to this model are shown (in bold) in 
the top panel of Fig.~\ref{fig:asca_rat}, along with the corresponding
ratios (dashed) for the data in the 0.40-0.60 keV and 
4.5-6.8 keV bands (observer-frame).
There are two points to note from Fig.~\ref{fig:asca_rat}: 
the indication of a steepening of the spectrum below
a rest-energy $\sim$0.7 keV and a marginal indication of line emission   
in the iron K$\alpha$ regime.

Given the suggestion of emission in the iron K$\alpha$ regime
we have repeated the 
spectral analysis, but including the data in the 4.5-6.8 keV band 
(observers-frame), we find the addition of a 
Gaussian emission component to the model  
provides a reduction in $\chi^2 $ of 11, for three fewer 
$dof$ (compared to 
a model without a Gaussian component).
This suggests the presence of an emission line 
at the 95\% confidence level. 
We find such a model to provide an adequate description 
of the data ($\chi^2=595/647\ dof$).
The line has a rest-energy $E=5.93^{+0.55}_{-0.40}$ keV, width 
$\sigma=0.26^{+0.74p}_{-0.26p}$ keV and intensity 
$I=2.4^{+2.9}_{-2.0} \times 10^{-5} {\rm photons\ cm}^{-2} 
{\rm s}^{-1}$ and equivalent width $EW=170^{+200}_{-150}$ eV. 
\footnote{A $p$ next to an error indicates that the error calculation 
reached the highest or lowest value allowed 
and thus the actual 90\% limit lies at or beyond the limit set in the 
fit for that parameter}
This is the first detection of iron K$\alpha$ emission from this source.  
While the line detection appears marginal in 
Fig.~\ref{fig:asca_rat}, we will show that the August data 
strongly confirm the existence of an iron K$\alpha$ line in this source. 
There were too few photons in the line to merit more detailed 
modeling of the profile.

\subsubsection{1996 August}

We treated these data in the same way as the July observation. 
Considering only the 2--10~keV band (rest-frame), but 
excluding the iron K$\alpha$ regime, we find a power-law attenuated by 
Galactic absorption provides an adequate description of the 
data ($\chi^2=378/389\ dof$) with
$\Gamma_{rest}$(2-10)$=1.61^+_-0.06$. 
Such a model also provides an adequate description of the 
full {\it ASCA} band (excluding the iron K$\alpha$ regime)
giving $\chi^2=635/642\ dof$ and 
$\Gamma_{rest}$(0.66-11.1)$=1.68^+_-0.03$, significantly steeper than 
the index observed in the July observation. 
The data/model ratio for this fit are shown (in bold)
in the bottom panel of Fig.~\ref{fig:asca_rat}. As above, the 
corresponding
ratios for the data in the 0.40-0.60 keV and 
4.5-6.8 keV bands (observer-frame) are shown dashed.
It can be seen that 
at this epoch there are also clear indications of a steepening of 
the spectrum below
a rest-energy $\sim$0.7 keV.
However, the {\it ASCA} data alone do not allow us to 
distinguish between the presence of a complex absorber, or 
of a separate continuum component. These data 
confirm that LB~1727 exhibits 
iron K$\alpha$ emission (Fig.~\ref{fig:asca_rat}). 

From repeating the spectral analysis, but including the data in the
iron K$\alpha$ band, we find the addition of a 
Gaussian emission component to the model is significant
at $>99$\% confidence (providing 
a reduction in $\chi^2 $ of 35 for three fewer $dof$ compared to 
a model without such a component).
As for the earlier epoch, such a model provides an adequate description 
of the data ($\chi^2=800/859\ dof$).
We find the line to have a rest-energy $E=6.39^{+0.68}_{-0.66}$ keV, 
with width $\sigma=1.0^{+0p}_{-0.35}$, intensity
$I=9.6^{+3.5}_{-5.4} \times 10^{-5}$ photons cm$^{-2}$ s$^{-1}$ 
and $EW=700^{+330}_{-400}$ eV at this epoch. 
Again, there were too few photons in the line to merit more detailed 
modeling of the profile. 
The photon index is unaffected by the addition of the line, and 
we note that the source appears steeper at this epoch, than during 
the July observation. with $\Delta \Gamma \sim 0.13$. 

\section{The ROSAT HRI Observations}

LB~1727 was monitored daily by the HRI, covering a two month period 
from 1996 June 30 to 1996 September 01, including a long integration 
simultaneous with the {\it ASCA} observation in July. 
The aim of the observations 
was to examine the variability behavior of this source on a number 
of timescales, particularly the days-to-weeks timescales 
which have not been extensively studied in Seyferts as a class. 

The HRI data were co-added and an image was produced. 
The HRI position for LB~1727 is 
(J2000) $04^{\rm h}26^{\rm m}0.9^{\rm s}\ -57^{\rm o}12'1.3''$, which 
agrees with the optical position to within $1''$. The image
shows that the only X-ray sources close to LB~1727 are relatively 
faint (a few per cent of the flux of LB~1727). 
The nearest two sources are $\sim 4'$ from the 
target and would be 
easily separable from LB~1727 in the SIS images if 
they had a significant hard X-ray flux. 

A radial profile was extracted and compared to the 
point-spread-function ($psf$) of the 
instrument. A small excess is observed over the $psf$, between 12-18$''$ 
from the centroid position. However, such an excess is commonly seen in
calibration (point) sources. This is thought to be a deficiency in the 
$psf$ model, and so we conclude there is no significant evidence for extended 
X-ray emission associated with LB~1727 in these data. 

Fig.~\ref{fig:hri_lc} shows a light curve based upon the {\it ROSAT} monitoring 
campaign. Source counts were extracted from a circular region of radius 
$30''$ encompassing 90\% of the source counts. 
The background level was small compared to the source flux. The 
background has not been subtracted from the source light curve 
but is shown on the same plot (rescaled to the 
level appropriate to the source extraction cell). 
We found no 
evidence for large amplitude and rapid variability in LB~1727. 
Rapid fluctuations at the 
few per cent level could be attributed to 
variations in the background rate. 
Significant variations in source flux were 
observed on timescales of $\sim 40$ ks. Moreover, the source flux 
increased by 
a factor of $\sim 3$ across the 2 month period covered 
by the HRI observations, 
one of the strongest increases being a factor of two change in 5 days,
close to the middle of the campaign. The amplitudes and timescales 
of variability seen in LB~1727 are 
similar to those observed in some other Seyfert 1 galaxies, like 
Mkn~335 (Turner \& Pounds 1988).  
The timescales are long and the amplitude of variability is small compared 
to behavior observed in some NLSy1 galaxies such as Mkn~478 (Marshall \etal\ 
1996) and IRAS 13224-3809, which has shown the most extreme short 
time variability in X-rays (Boller \etal\ 1997).

\section{Analysis of the Simultaneous HRI and ASCA data}

The known strength of the XUV flux (e.g. Marshall \etal\ 1995) 
suggests we might expect to observe a strong spectral break between the 
soft and hard X-ray regimes in this source. There is an indication 
that this does indeed occur, based on the 
turn-up observed in the lowest channels of the SIS data 
(Fig.~\ref{fig:asca_rat}). 
We fit the simultaneous HRI and {\it ASCA} data together, and found 
that a simple extrapolation of the {\it ASCA} model significantly 
underpredicted the HRI flux (Fig.~\ref{fig:hri_asca}). 
The key question then becomes, could the presence of ionized gas, 
the so-called ''warm-absorber'', 
be the cause of the observed spectral steepening, or is that primarily due to 
a steepening of the continuum form. 

To answer this question we fit the data using 
models based upon the photoionization code ION (Netzer 1996). Here we use the 
ionization parameter, $U_X$, where 
\begin{equation}
U_X = \int^{\nu (10\ {\rm keV})}_{\nu(0.1\ {\rm keV})}
\frac{L_{\nu} / h \nu}{4 \pi r^2 n_{H} c} d\nu
\end{equation}
$L_{\nu}$ is the monochromatic luminosity, $r$ the distance from the
source to the illuminated gas. 
$U_X$ provides a quantity directly proportional to 
the level of ionization of the dominant species observed in the X-ray regime. 
Comparison between $U_X$, and the other ionization parameters  
can be found in George \etal\ (1998), where the ION model was applied to a 
sample of Seyfert~1 galaxies. 
It was not possible to find an acceptable fit using a single 
power-law and an ionized absorber when the HRI point was included in 
the fit (the best-fit yielded $\chi^2=636/532\ dof$ which is unacceptable 
at $>$ 99\% confidence). 
However, the simultaneous data are adequately fit with a 
broken power-law model, which gives 
$\Gamma_{hard}=1.45^+_-0.05$, $\Gamma_{soft}=3.63^{+6.37p}_{-0.46}$, 
a break energy of $0.75^{+0.03}_{-0.36}$ keV   
and $\chi^2=488/532\ dof$. Thus we conclude that a steepening 
of the continuum is the most natural explanation for the spectral 
break, and complex absorbers alone cannot mimic such a sharp 
up-turn. 

Despite the requirement for a steepening continuum, 
we cannot {\it rule out} the presence of some ionized gas. 
If the source has a steepening continuum {\it and} is also 
modified by transmission through a column of (unresolved) ionized gas, 
then the observed break is a combination of two effects, a 
turn-up of the emission spectrum and a reduction in gas opacity 
at soft X-ray energies. We find the addition of a warm-absorber 
does not improve the fit at all, and if the gas 
is very highly ionized  we cannot constrain the 
amount of gas which might be present. If the ionization 
parameter is constrained such that $U_X < 10$, as it was 
for the Seyfert 1 analysis (George \etal\ 1998) then 
we find 
$ 0 < N_H^{*} < 2.42 \times 10^{22} {\rm cm}^{-2}$ 
(90\% confidence) with a lower limit $U_X > 0.42$. 
It is evident that large amounts of highly-ionized gas could be 
present but undetectable. 

The data are inconsistent with the screen of neutral gas 
of column $2 \times 10^{21} {\rm cm}^{-2}$ estimated assuming the 
Balmer decrement is an indicator of extinction, 
even if the covering fraction of the gas is allowed 
to be as small as 10\%.

\section{The PSPC dataset} 

The {\it ROSAT} PSPC data are of interest with respect to the 
determination of the soft X-ray spectrum, in particular the 
energy of the spectral break. 
LB~1727 was observed by the {\it ROSAT} PSPC on 1992 April 07. 
Analysis of these PSPC data has been presented by Guainazzi \etal\ (1998). 
Here we present our independent analysis, which is generally consistent 
with that presented by Guainazzi \etal\ (1998). The data were 
corrected for time-dependent effects using the ftool 
{\sc pcpicor}.  Source data were extracted 
from a cell of radius $3'$ and  background data from an 
annular region centered on the source. The source flux was 
$F_{rest}$(0.5-2 keV)$=1.9 \times 10^{-11} {\rm erg\ cm}^{-2} {\rm s}^{-1}$ 
corresponding to a luminosity 
$L$ (0.1-2 keV) $\sim 10^{45} {\rm erg\ s}^{-1}$. 
This flux is approximately a factor of 4 brighter than that measured by 
{\it ASCA} in the same bandpass 4 years later and 
increased by only $\sim 5\%$ over the 40 ks separating the 
observation intervals comprising this dataset.  

The mean spectrum was fit with a simple power-law, attenuated by a 
neutral absorber, which was unconstrained. It 
was impossible to achieve an acceptable fit 
using this model. The data/model ratio plot is shown in 
Fig.~\ref{fig:rosat_rat}, 
demonstrating a sharp break in the spectrum, confirming the 
result from the {\it ASCA} and HRI data. 
Application of a broken power-law model yields a 
rest-energy $0.70^+_-0.08$ keV for the spectral break. 
The photon index above 0.7 keV is 
$\Gamma=2.61^+_-0.15$. The data below the 
break energy are not well fit with a simple power-law. In this regime, 
much of the contribution to $\chi^2$ arises 
from a mixture of positive and negative data-minus-model residuals in the 
0.3-0.4 keV regime. These may indicate the presence of emission and/or 
absorption features. However, 
this bandpass is where the effective area of the instrument has a steep 
gradient and modelling of features in this regime is 
strongly subject to any small residual inaccuracies in the area curve. 
{\it Parameterizing} this part of the spectrum 
with a power-law gives a photon index 
$\Gamma=3.72^{+1.18}_{-0.33}$ below 0.7 keV. In this case 
the fitted column density was 
$N_H=2.93^{+0.70}_{-0.22} \times 10^{20} {\rm cm}^{-2}$, 
slightly higher than the Galactic value and yielding 
$\chi^2=35/15\ dof$. If the neutral column density is fixed
at the Galactic value then the soft index is 
found to be $\Gamma=3.24^+_-0.08$, but the fit is worse with 
$\chi^2=51/16\ dof$. 
Analysis of the {\it ROSAT} All-Sky Survey data (RASS) 
spectrum confirms the presence of a spectral break at 
$\sim 0.7$ keV, 
and shows spectral slopes consistent with the pointed observation. 

An alternative model to describe the spectral shape is that of 
a power-law with an absorption edge. Such a fit 
yields $\chi^2=24/14\ dof$ for photon index 
$\Gamma=3.02^{+0.14}_{-0.11}$ and a rest-energy 
$E=0.58^+_-0.05$ keV for an edge of depth $\tau=0.96^{+0.34}_{-0.32}$. The 
column density was 
$N_H=2.50^{+0.40}_{-0.34} \times 10^{20} {\rm cm}^{-2}$. Again, the 
dominant contributions to $\chi^2$ are close $\sim 0.3$ keV. In this case 
the underlying continuum is steeper than that in the {\it ASCA} regime. 
The fit featuring an absorption edge is a superior description 
of the PSPC data, but there is no evidence for a deep absorption feature in 
the 0.4-0.6 keV SIS data (overlaid in Fig.~\ref{fig:asca_rat}). 
We find that the combined HRI and {\it ASCA} data cannot be 
described by a single power-law plus a deep absorption feature 
(notch or edge), and an absorption feature is not required once we 
allow the data to be fit with a broken power-law. 

The pointed observation from the PSPC 
found the source to be much brighter, and the 
0.6-2.0 keV index much steeper than during the 
{\it ASCA} epochs. Also, for all of the aforementioned fits, the PSPC slope is 
inconsistent with the {\it ASCA} spectral index in the overlapping  bandpass 
(which is effectively 0.6-2.0 keV). This disagreement is either 
attributable to spectral variability of the source, or
to a greater degree of inconsistency in the cross-calibration 
between the PSPC and {\it ASCA} than previously thought. An astrophysical 
explanation might be that the soft spectral component was relatively strong 
during the PSPC epoch, and dominated the spectrum up to a higher energy 
than during the {\it ASCA} observation. 
In this case one would also expect the break to move to a higher energy,
which is not observed.

\section{The IR to X-ray Spectrum}

In Fig.~7 we examine the shape of the SED by utilizing 
the multiwaveband data compiled by Grupe et al. (1998a) and combining our 
{\it ASCA} and {\it ROSAT} data with 
infra-red, optical fluxes and RASS data. All data are 
 corrected for known sources of absorption 
and the {\it EUVE} data represent the  
average flux based upon data from 
 1992 July through 1993 July (Marshall \etal\ 1995; 
Fruscione 1996). The {\it IUE} data  
were extracted from the archival observation of 1994 Oct 27. 
The {\it IUE} spectrum shows strong emission lines from Ly$\alpha$ and
C{\sc iv} (at z=0.104) with no evidence for absorption. 
The {\it IUE} data lie a little low compared to the {\it EUVE} data, 
however these are not simultaneous with any other dataset and 
large amplitude variations are a property of the XUV-bump (Fig.~4). 

{\it ASCA} and PSPC data are represented by 
pseudo-bow-ties showing the 90\% confidence 
ranges of spectral index and folding in the (10\%) uncertainty in 
the absolute flux calibration. Table \ref{slopes} 
summarizes the spectral slopes of  LB~1727 between 100 $\mu$m and 1 keV, 
for comparison with those calculated for a large sample of AGN as detailed 
in Grupe \etal\ (1998a).  
The index between 2500\AA\ and 2 keV is also calculated, as this 
is widely quoted in the literature. 
We utilized the PSPC spectrum from the pointed observation of 
{\it ROSAT} to obtain 
0.25, 1 and 2 keV flux points for index determination. The pointed data 
allow a more accurate determination of the fluxes than the RASS data, 
and over a softer bandpass than the {\it ASCA} data. 
LB~1727 lies within the range of indices 
found between 5500\AA\ and 0.25, 1 keV for Seyfert galaxies 
(Grupe \etal\ 1998a) but at the extreme 
end of the range, indicating the source to be relatively 
bright in the soft X-ray band.  However, we note that LB~1727 
was relatively bright during the pointed PSPC observation, and 
the derived indices are slightly shallower than they would 
be if we had used the RASS or {\it ASCA}  
data. Obviously, combining non-simultaneous data can only give an 
approximation to the SED, for such a variable source. 
The index between 2500\AA\ and 2 keV is 1.23, if we assume a 2 keV flux 
a factor of four dimmer than observed during the PSPC observation 
(as observed by {\it ASCA}), then the index would be 
1.46. These values lie within the normal range for a Seyfert galaxy 
(Kriss \& Canizares 1985).

\begin{table}[ht]
\caption {\label{slopes} Spectral slopes of  LB~1727 in the IR to X-ray
range
} 
\begin{center}
\begin{tabular}{lcl}
\noalign{\smallskip} \hline \noalign{\smallskip}
Quantity  & Slope & Definition \\
\noalign{\smallskip} \hline \noalign{\smallskip}
$\alpha_{\rm 7000 \AA\ - 4400 \AA\ }$&0.01&-4.967 
log$\frac{f_{4400\AA\ }}{f_{7000\AA\ } }$ \\
\noalign{\smallskip}
$\alpha_{\rm 5500\AA\ - 0.25\ keV}$ &0.73& -0.489 log$\frac{f_{0.25\ keV }}{f_{5500\AA\ }}$ \\
\noalign{\smallskip}
$\alpha_{\rm 5500\AA\ - 1\ keV}$ & 1.13&-0.378 log$\frac{f_{1\ keV}}{f_{5500\AA\ } }$ \\
\noalign{\smallskip}
$\alpha_{\rm 60\mu-5500\AA\ }$ & 0.70&-0.491 log$\frac{f_{5500\AA\ }}{f_{60\mu} } $ \\
\noalign{\smallskip}
$\alpha_{\rm 60\mu-0.25\ keV}$ & 0.72 & -0.245 log$\frac{f_{0.25\ keV} }{f_{60\mu} }$  \\
\noalign{\smallskip}
$\alpha_{\rm 60\mu-1\ keV}$ & 0.94 & -0.214 log$\frac{f_{1\ keV}}{f_{60\mu}} $  \\
\noalign{\smallskip}
$\alpha_{\rm 12\mu-5500\AA\ }$ & 0.84 & -0.747 log$\frac{f_{5500\AA\ }}{f_{12\mu}} $ \\
\noalign{\smallskip}
$\alpha_{\rm 12\mu-0.25\ keV}$ & 0.78&-0.296 log$\frac{f_{0.25\ keV} }{f_{12\mu} } $ \\
\noalign{\smallskip}
$\alpha_{\rm 12\mu-1\ keV}$ & 1.04&-0.251 log$\frac{f_{1\ keV} }{f_{12\mu} } $ \\
\noalign{\smallskip}
$\alpha_{\rm 2500\AA\ - 2\ keV}$ & 1.23&-0.384 log$\frac{f_{2\ keV} }{f_{2500\AA\ } } $ \\
\noalign{\smallskip}\hline\noalign{\smallskip}
\end{tabular}
\end{center}
\end{table}

As mentioned in section \ref{sec:opt}, the soft X-ray 
and UV properties of LB~1727 might lead us to expect it to fall
into the subclass of NLSy1s, but it does not; it has a broad 
width for FWHM(H$\beta$), 
the \verb+[+O{\sc iii}\verb+]+/H$\beta$ ratio is high and 
there is little evidence for Fe{\sc ii} emission. 
We investigated whether  LB~1727 fits into the normal 
correlations found between parameters in other 
AGN (e.g. Boroson \& Green 1992, Grupe et al. 1998b). 
We found LB~1727 to have a relatively large ratio 
\verb+[+O{\sc iii}\verb+]+/H$\beta$ for a source with such a 
flat optical continuum slope (Grupe et al. 1998b). 
Using the value of X-ray index based upon a single power-law fit to the 
RASS data (Grupe \etal\ 1998a) we find LB~1727 fits into the  
$\alpha_{\rm X}$ - FWHM(H$\beta$) 
relationship  observed for Seyfert galaxies (Boller \etal\ 1996). 
If we were to use the steeper spectral index observed below 0.75 keV 
then this would no longer hold true.  However, using parameters 
from the broken power-law fit would no longer 
yield a fair comparison with the simple indices plotted for other sources 
in those samples, which were fit with a single power-law model. 

Compared to both hard and soft X-ray selected samples of AGN, we find 
LB~1727 to be relatively weak in the IR regime (cf. Grupe \etal\ 1998a).
This is also evident by comparison of the SED with that of other 
hard and soft X-ray selected AGN (Fig.~4 of by Grupe \etal\ 1998a). 
This suggests that if dust exists in this source, it is not 
excited by the nuclear radiation.

\section{Discussion} 

The hard X-ray spectrum of LB~1727 is flat, 
with $\Gamma=1.5-1.6$ between rest-energies $\sim 0.75-11$ keV, and 
shows no evidence for intrinsic absorption by neutral or ionized material.
There are several examples of flat X-ray (2-10 keV) spectra reported for 
Seyfert galaxies (NGC 4151, e.g. Weaver \etal\ 1994; 
NGC 3227, e.g. George \etal\ 1998, Mrk~6, George 1999); however, in most known 
cases these nuclei are also heavily absorbed. 
The observation of a flat continuum 
does not appear to be an artifact of confusion between continuum 
and Compton reflection, or between continuum and a complex 
absorber. We  added a 
model component representing the hard spectral feature due to Compton 
reflection from neutral material 
(Magdziarz \& Zdziarski,  1995). 
We found the July 1996 {\it ASCA} data 
still required a flat underlying continuum, $\Gamma=1.45^+_-0.06$, the 
additional component did not improve the fit at all, and gave an upper limit 
(at 90\% confidence) 
on the solid angle of the reflector, $\Omega/2\pi < 2$ (for reflection 
from a disk of material observed face-on). 
For the August 1996 {\it ASCA}  data addition of the same model component 
resulted in a reduction  $\Delta \chi^2=5$, and gave a solution with
$\Omega/2\pi=1.7^{+2.0}_{-1.6}$. In that case the inferred underlying index 
steepened to $\Gamma=1.74^+_-0.05$. Thus the {\it ASCA} data are inconclusive 
on the importance of Compton reflection in this source. Models assuming 
reflection from ionized material may prove appropriate, however, given 
the limited bandpass of {\it ASCA} these 
data do not merit any more detailed modeling related to reflection. 

Allowing the presence of large amounts of absorbing material, 
ionized or neutral, fully or partially-covering the source,  did 
not reveal a solution consistent with a significantly steeper 
continuum slope. No physical explanation  has been offered  
linking spectral index and absorption. It is, however, 
plausible that the ionizing continuum could influence the 
condition of the line-of-sight gas and vice versa. 
In light of such possibilities it is 
interesting to find examples of flat, but apparently unabsorbed 
spectra. 
The source also shows line emission from the K-shell of iron. Unfortunately 
there are too 
few line photons to constrain the ionization-state of the emitting material, 
or the width of the line.

The requirement for a marked steepening of the soft X-ray continuum, 
combined with a rise in the optical continuum bluewards of 
$\sim 4500$\AA\ 
indicates that a spectral component exists which peaks between these two 
extremes, the so-called XUV-bump. While this has been inferred for many 
Seyfert galaxies in the past (starting with Arnaud \etal\ 1985), 
the discovery of the existence of large amounts of ionized gas 
along the line-of-sight to many Seyfert nuclei (e.g. Nandra \& Pounds 1992,  
Reynolds 1997, George \etal\ 1998) confused the issue. 
In many sources, it has become unclear whether the observed 
steepening to soft X-ray energies 
is predominantly due to the reduced opacity of ionized species below 
$\sim 1$ keV in an ionized absorber. Thus 
the existence of an XUV-bump has been called into question 
(e.g. Nandra \etal\ 1995, Laor \etal\ 1997). Examination of LB~1727 shows 
that in this case, the XUV-bump is required, even if we allow the 
possibility of some attenuation by unresolved, ionized gas. 
Assuming the nucleus is unattenuated then the spectral break 
is constrained (by the combined HRI and {\it ASCA} data) to lie at 
$0.75^{+0.03}_{-0.36}$ keV. If ionized material is present, then, while
we know that this cannot produce all of the observed spectral steepening it
may contribute to it. In this case the {\it continuum} break can only be 
constrained to lie at a rest-energy in the $\sim 0.13$ -- 0.75 keV range 
(the lower limit being given by the effective rest-energy of the 
lower bound of the HRI bandpass). 

While LB~1727 has spectral indices which are extreme, such 
properties are not unknown. The source resembles 
the Seyfert~1 galaxy  Mrk~841, which 
shows a two-component X-ray spectrum. The index of Mrk~841 is 
variable in the 2 -- 10 keV band and has been observed to be 
as flat as that reported here for LB~1727 (George \etal\ 1993). 
The spectrum of Mkn~841 
may also steepen to lower energies, this has been suggested to be due to the 
presence of an XUV-bump, although, as noted above, recent analysis
 shows the shape of the soft X-ray spectrum 
could be attributable to the effects of ionized gas along the line-of-sight 
in that case, and the soft X-ray spectrum can 
be simply extrapolated to meet the UV data (Nandra \etal\ 1995). 

In the case of LB~1727, the XUV-bump must be a source of 
copious ionizing photons. Assuming there is no absorption 
between this bump component and the optical-line-emitting regions then 
it should 
have a noticible effect on optical line ratios. As discussed 
by Cohen (1983) and Kraemer \etal\ (1999), the 
He{\sc ii}$\lambda4686$/H$\beta$ ratio should depend strongly 
on the XUV spectrum. In general, examination of the narrow components 
of these lines is illuminating, as the narrow-line-region 
is free from the strong collisional effects affecting the 
broad-line-region. As shown by Kraemer \etal\ (1999), LB~1727 has a 
relatively strong He{\sc ii} 
contribution which may be linked to the presence of a strong XUV-bump. 

The Balmer decrement determined from the sum of the broad and narrow 
components of H$\alpha$ and H$\beta$ 
could be due to the presence of a column density 
$N_H\sim 2 \times 10^{21} {\rm cm}^{-2}$ along the line-of-sight. 
A column of ionized, 
dusty material would be consistent with data in both the optical and X-ray 
regimes. However, 
the SED indicates that dust is not important in this source and in any case 
some simpler explanations are more compelling. The lack of attenuation 
of the optical continuum suggests the Balmer decrement is due to opacity 
effects in the line-emitting clouds. Alternatively,
the gas which obscures the regions producing optical lines may simply be 
out of the line-of-sight to the nucleus. 

While the HRI and {\it ASCA} data have allowed us to determine that the 
continuum breaks to a steeper form at $\sim 0.75$ keV, we can do little to 
examine the detailed spectrum of the soft component. Fortunately it 
will soon be possible to examine this interesting source with {\it AXAF} 
and {\it XMM}. 
The high spectral resolution afforded by the {\it AXAF} 
and {\it XMM} gratings, along 
with their broad-bandpasses, extending down to the soft X-ray regime, 
allows the 
opportunity for a significant advance in our understanding of LB~1727.

\section{Acknowledgements}
We are grateful to \asca\ team for their operation of the satellite, 
to Hagai Netzer for use of ION 
and to Jules Halpern for very useful comments. This research has 
made use of the NASA/IPAC Extragalactic database,
which is operated by the Jet Propulsion Laboratory, Caltech, under
contract with NASA; of the Simbad database, 
operated at CDS, Strasbourg, France; and data obtained through the 
HEASARC on-line service, provided by NASA/GSFC. We acknowledge the 
financial support 
of Universities Space Research Association (IMG, TJT) and 
the National Research Council (KN).

\newpage

\clearpage
\pagestyle{empty}
\setcounter{figure}{0}
\begin{figure}
\plotfiddle{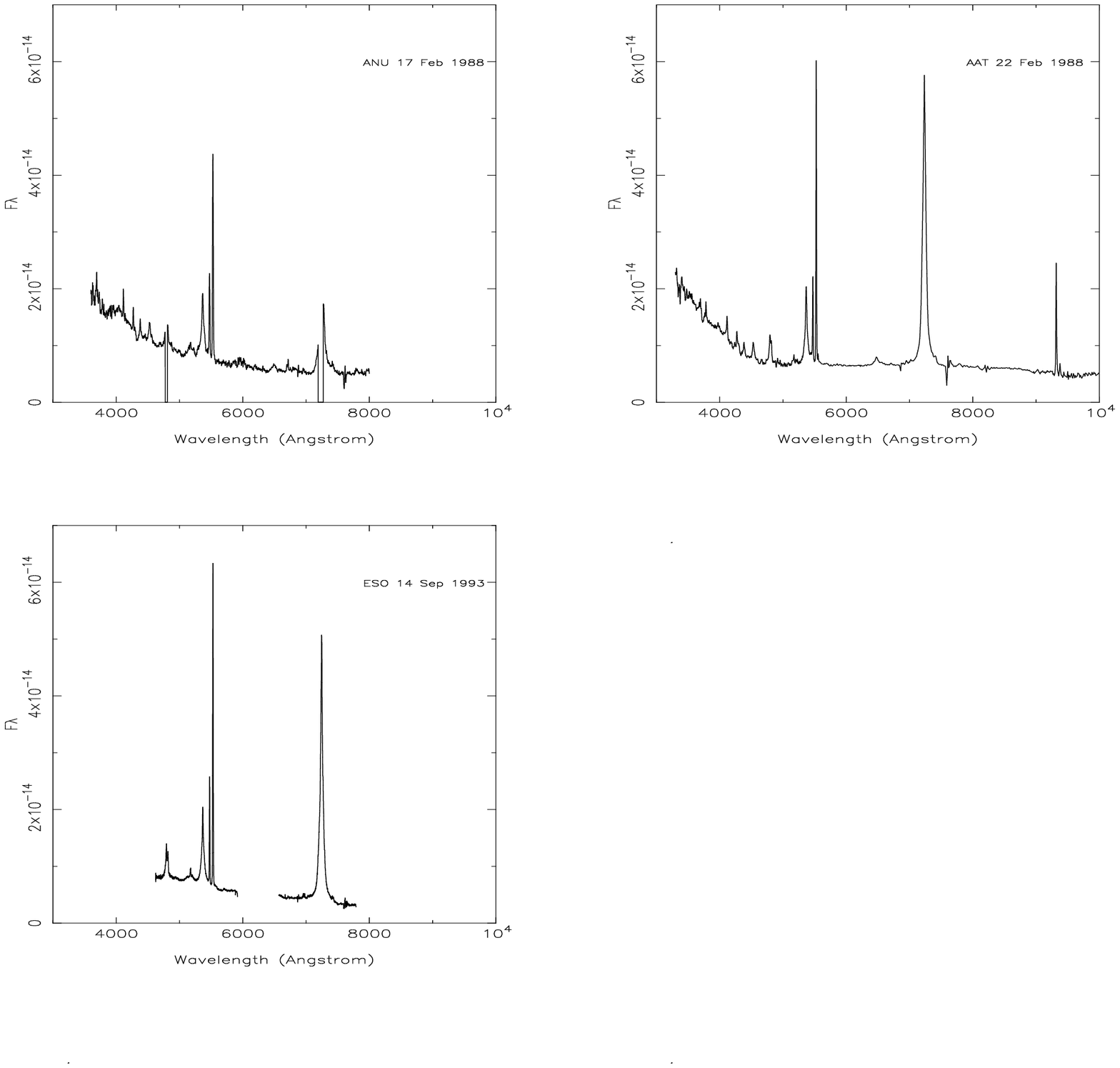}{10cm}{0}{50}{0}{-200}{0}
\caption{
\label{fig:opt}
Optical spectra of LB~1727 taken with the 
Australian National University (ANU) 2.3m 
and the MPI/ESO 2.2m telescopes and 
3.9m Anglo-Australian Telescope (AAT). In the case of the ANU data 
the red and blue portions of the spectra were each measured with
a pair of photon-counting arrays, and the spaces between these
detectors causes the $\sim$ 5 pixel gaps seen in the spectrum.
The ANU and AAT data clearly illustrate a steep rise in the 
continuum blueward of 4500 \AA.
}
\end{figure}
\clearpage

\setcounter{figure}{1}
\begin{figure}
\plotfiddle{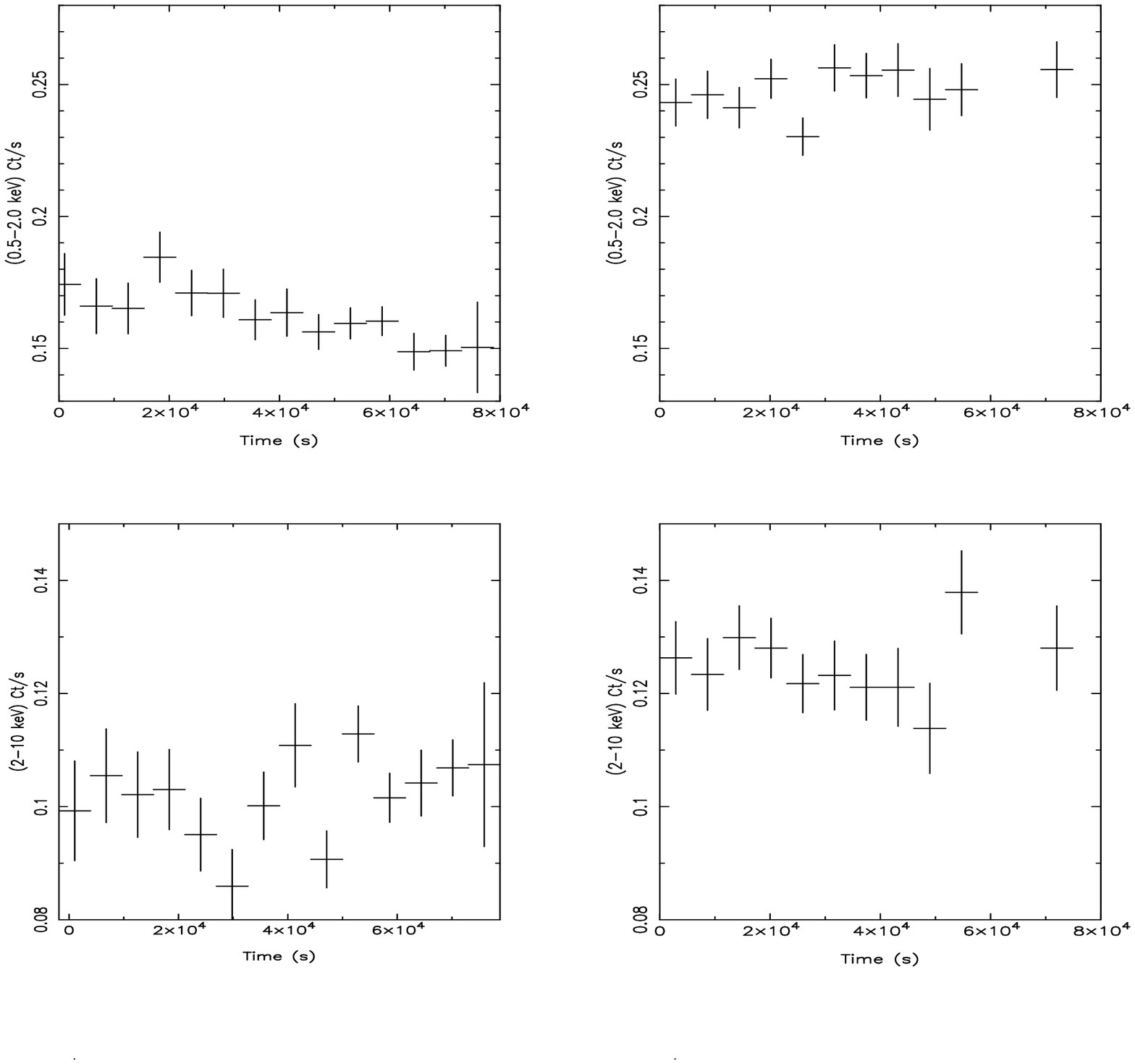}{10cm}{0}{50}{50}{-200}{0}
\caption{
\label{fig:asca_lc} 
The light curves in 5760 s bins for the combined SIS data in the 
observed-frame 
0.5-2 keV band (upper panel) and 2-10 keV band 
(lower panel) for the 1996 July (left panels) 
and 1996 August (right panels)
data.}
\end{figure} 
\clearpage

\setcounter{figure}{2}
\begin{figure}
\plotfiddle{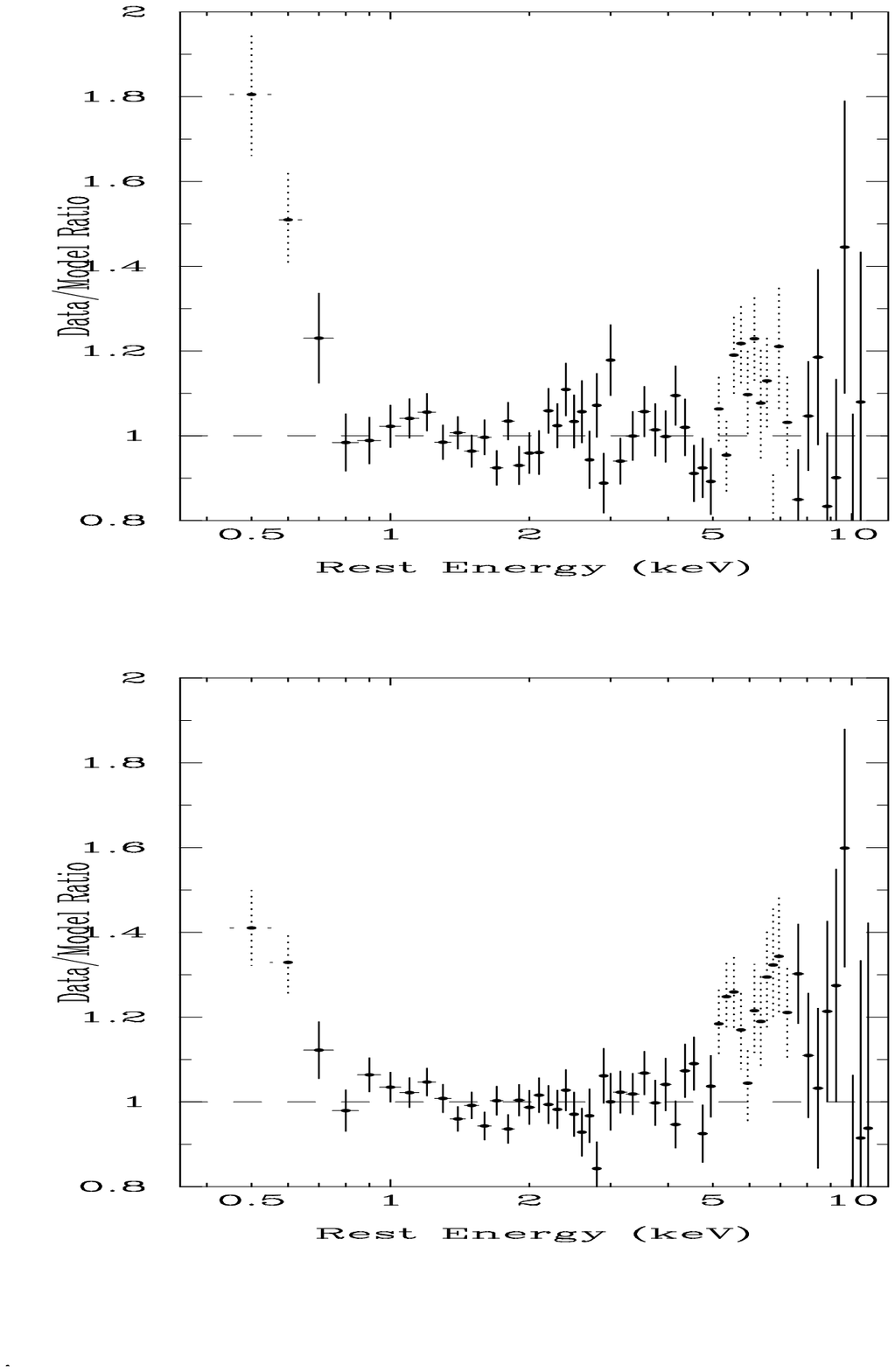}{10cm}{0}{50}{50}{-200}{0}
\caption{
\label{fig:asca_rat}
The data/model ratio from the combined SIS+GIS data, 
compared to a power-law model.  The dotted 
points show the SIS data from the 0.4-0.6 keV band, which were not used 
in the fit but have been overlaid for illustrative purposes. Data are 
compared to a power-law model with absorption by a column of neutral material, 
fixed at the Galactic line-of-sight value.
a) July 1996 data with $\Gamma=1.45$ b) August 1996 data with $\Gamma=1.68$ 
}
\end{figure}
\clearpage

\setcounter{figure}{3}
\begin{figure}
\plotfiddle{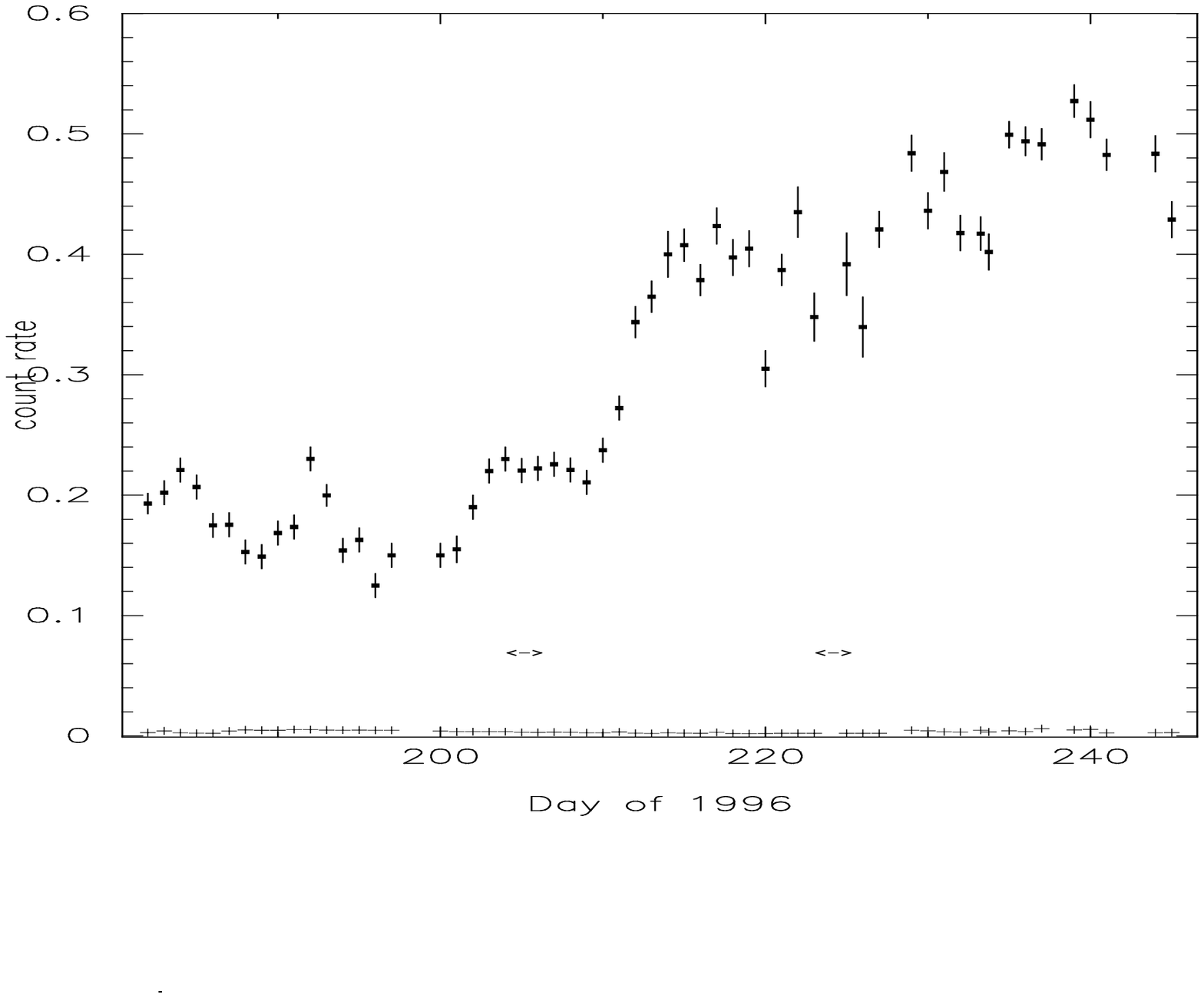}{10cm}{0}{50}{0}{-200}{0}
\caption{
\label{fig:hri_lc}
The HRI light curve, showing the source count rate taken from a 
cell of $30''$ radius which encompasses 90\% of the source counts. 
The background 
level in the source cell is shown as the lower light curve and is
$< 10^{-2} {\rm ct\ s}^{-1}$ throughout. 
The two small double-headed arrows show 
the periods when {\it ASCA} was observing the source.
}
\end{figure}

\clearpage
\setcounter{figure}{4}
\begin{figure}
\plotfiddle{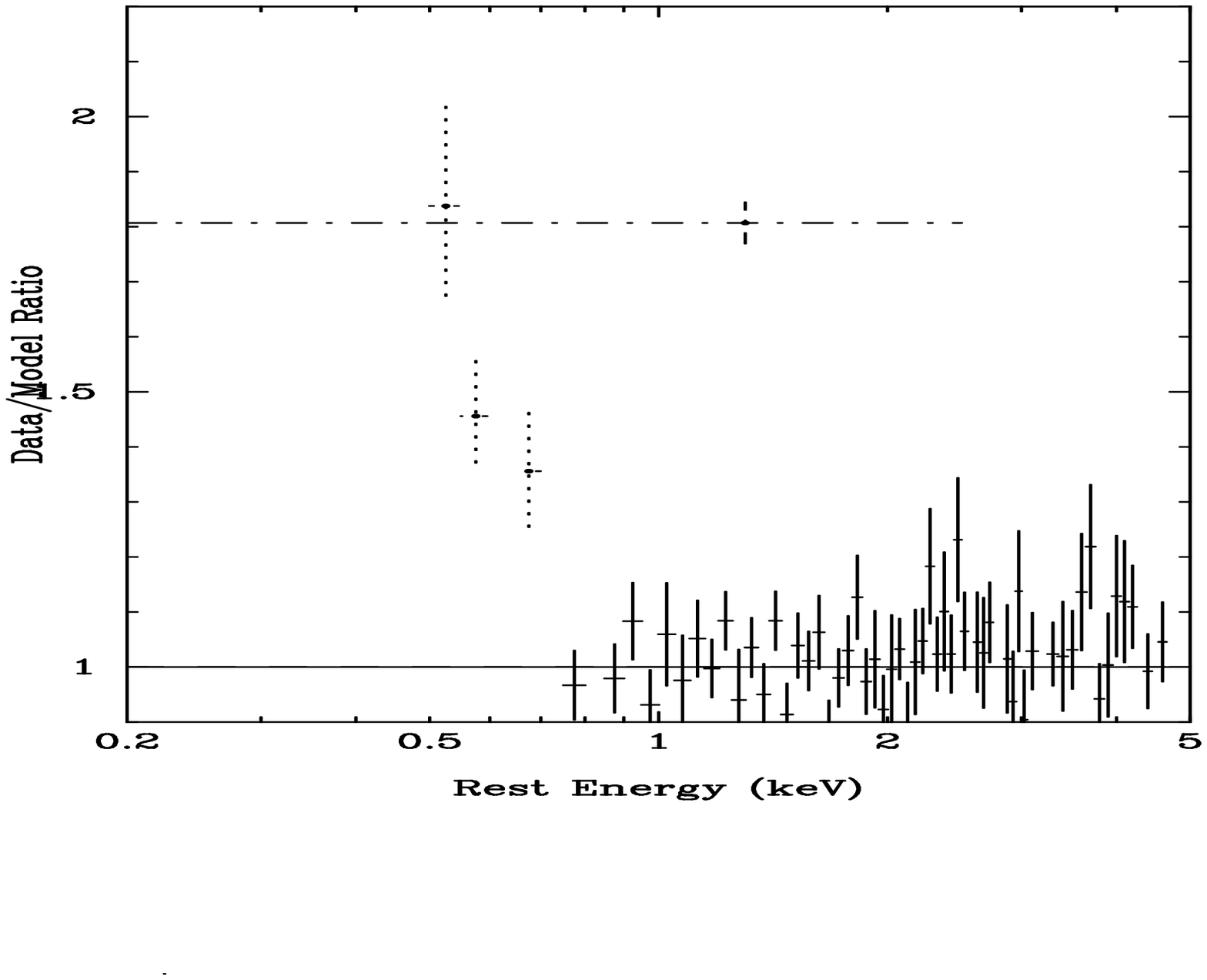}{10cm}{0}{50}{0}{-200}{0}
\caption{
\label{fig:hri_asca}
The ratio of the 1996 July {\it ASCA} data and simultaneous HRI data to a model 
of a power-law attenuated by an ionized absorber, after fitting the 
{\it ASCA} data (above 0.6 keV). The {\it ASCA} data below an observed-frame 
energy of 0.6 keV are shown as dotted lines, as these were not used in the fit. 
The HRI data are shown as a dashed point. Fitting this HRI point 
simultaneously with the {\it ASCA} data did not yield an acceptable fit 
and indicates the presence of a spectral break at a rest-energy 
$\sim 0.75$ keV, (see text for details). 
} 
\end{figure}

\setcounter{figure}{5}
\begin{figure}
\plotfiddle{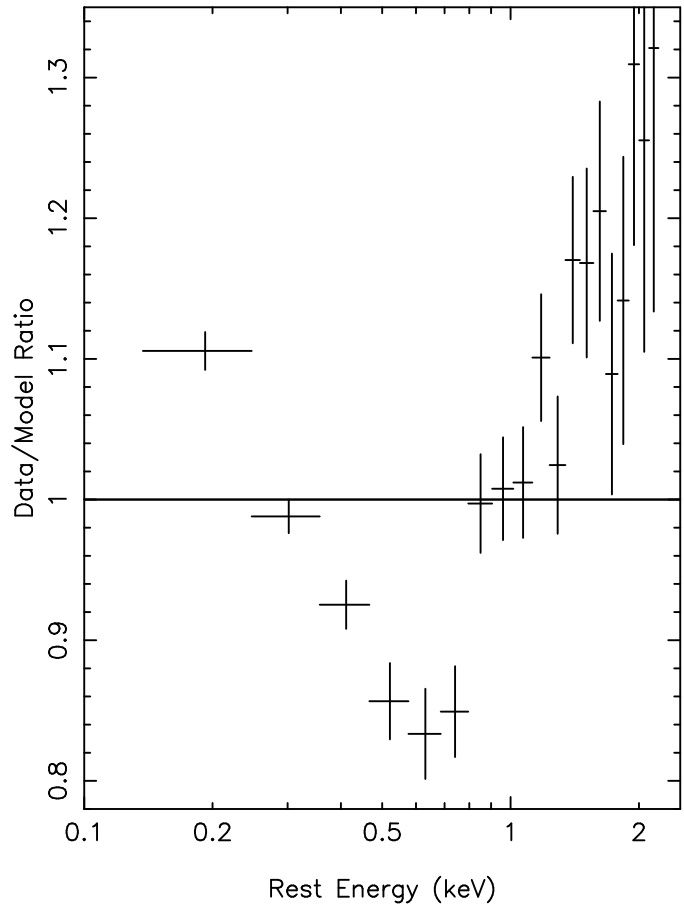}{10cm}{0}{50}{50}{-200}{0}
\caption{
\label{fig:rosat_rat}
The data/model ratio from the {\it ROSAT} PSPC data, compared to a power-law
model, allowing attenuation by an unconstrained column of neutral material,
showing the spectral break at a rest-frame energy of $\sim 0.75$ keV.
}
\end{figure}
\clearpage

\clearpage
\clearpage
\setcounter{figure}{6}
\begin{figure}
\plotfiddle{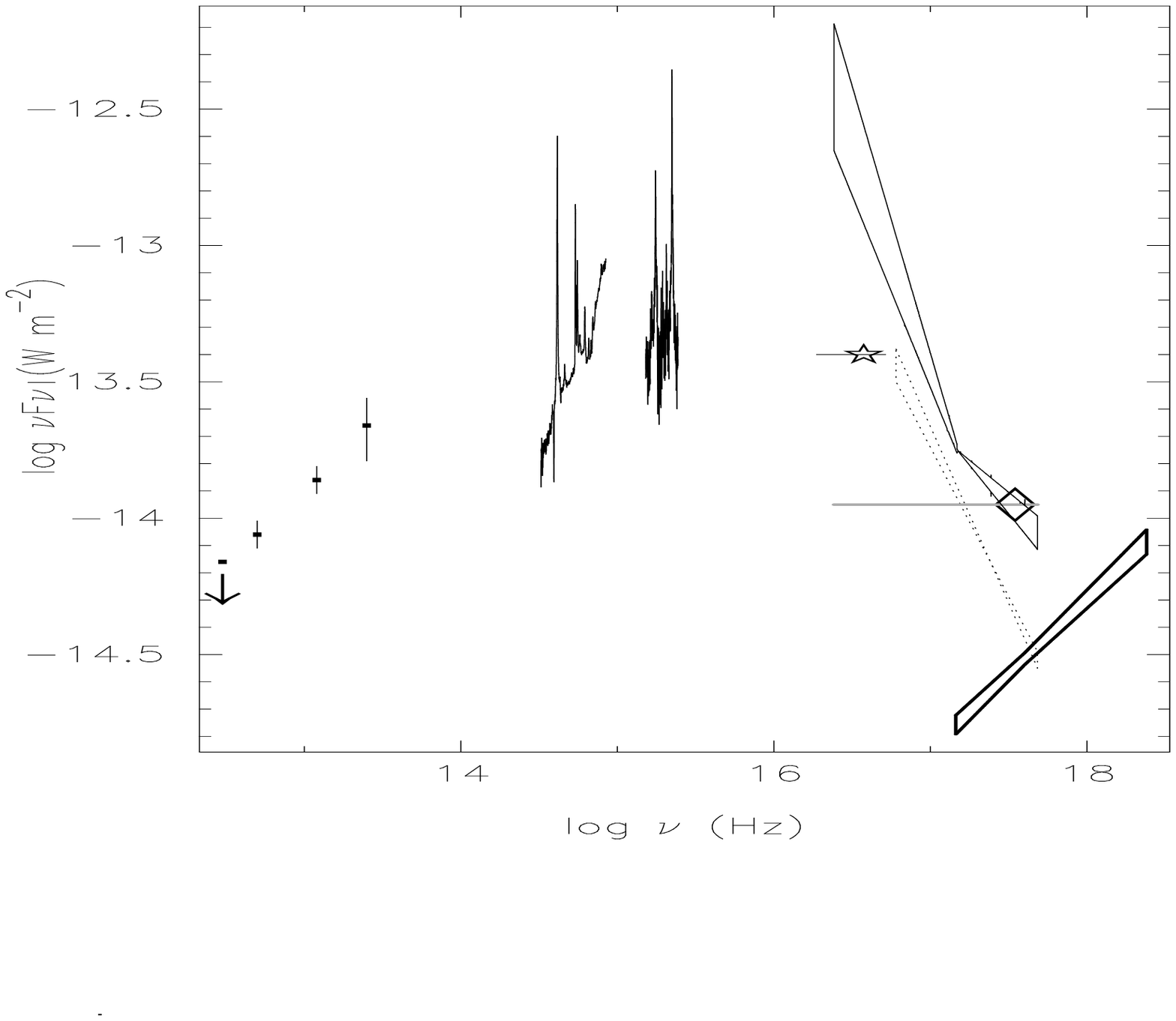}{10cm}{0}{50}{0}{-200}{0}
\caption{
\label{fig:mwb}
The (pseudo) bow-tie marked by the thick solid line represents 
the {\it ASCA} spectrum for 
July 1996, along with the simultaneous HRI point (diamond with 
horizontal line showing the bandpass), corrected for Galactic absorption. 
Also shown are non-simultaneous 
multiwaveband data as compiled by Grupe (1996);  
{\it EUVE} data (open star); 
{\it ROSAT} PSPC pointed data (solid bow-tie line); 
RASS data (dotted bow-tie); IR fluxes 
(solid squares with the upper limit shown for the 
100$\mu$ flux), and {\it IUE} spectrum 
 and a low resolution version of the optical spectrum 
from the ESO 2.2m telescope. 
} 
\end{figure}

\end{document}